\documentclass[sigconf]{acmart}

\AtBeginDocument{%
	\providecommand\BibTeX{{%
			\normalfont B\kern-0.5em{\scshape i\kern-0.25em b}\kern-0.8em\TeX}}}

\usepackage{amsmath,amssymb,amsfonts}
\usepackage{algorithmic}
\usepackage{graphicx}
\usepackage{textcomp}
\usepackage{subfigure}
\usepackage{multirow}
\usepackage{booktabs}
\usepackage{epstopdf}
\usepackage{color}
\usepackage{algorithm}
\begin{document}
	
\title{GAN-Based Multi-View Video Coding with Spatio-Temporal EPI Reconstruction}
\author{Chengdong Lan}
\email{lancd@fzu.edu.cn}
\affiliation{%
	\institution{Fuzhou University}
	\country{}}

\author{Hao Yan}
\email{how\_lin@163.com}
\affiliation{%
	\institution{Fuzhou University}
	\country{}}

\author{Cheng Luo}
\email{n191120073@fzu.edu.cn}
\affiliation{%
	\institution{Fuzhou University}
	\country{}}

\author{Tiesong Zhao}
\email{t.zhao@fzu.edu.cn}
\affiliation{%
	\institution{Fuzhou University}
	\country{}}

\renewcommand{\shortauthors}{}

\begin{abstract}
The introduction of multiple viewpoints in video scenes inevitably increases the bitrates required for storage and transmission. To reduce bitrates, researchers have developed methods to skip intermediate viewpoints during compression and delivery, and ultimately reconstruct them using Side Information (SI). Typically, depth maps are used to construct SI. However, their methods suffer from inaccuracies in reconstruction and inherently high bitrates. In this paper, we propose a novel multi-view video coding method that leverages the image generation capabilities of Generative Adversarial Network (GAN) to improve the reconstruction accuracy of SI. Additionally, we consider incorporating information from adjacent temporal and spatial viewpoints to further reduce SI redundancy. At the encoder, we construct a spatio-temporal Epipolar Plane Image (EPI) and further utilize a convolutional network to extract the latent code of a GAN as SI. At the decoder side, we combine the SI and adjacent viewpoints to reconstruct intermediate views using the GAN generator. Specifically, we establish a joint encoder constraint for reconstruction cost and SI entropy to achieve an optimal trade-off between reconstruction quality and bitrates overhead. Experiments demonstrate significantly improved Rate-Distortion (RD) performance compared with state-of-the-art methods.
\end{abstract}


\renewcommand\footnotetextcopyrightpermission[1]{}
\settopmatter{printacmref=false}

\maketitle

\section{Introduction}
To provide more immersive experience, multi-view video captures visual information from different positions and angles, and thereby leading a surge in the amount of data. How to reduce the coding bitrates while ensuring the reconstruction quality has become a critical issue. Recent efforts have confirmed the feasibility of deep learning-based video coding \cite{ Hu_2021_CVPR, liu2020deep, Li:NIPS21, Zhao_2022, Zhang:IS2020}. This is benefited from the training of large datasets coupled with the powerful nonlinear modeling capability of neural networks. Unfortunately, little research has been conducted on deep learning-based Multi-view Video Coding (MVC) \cite{tech2015overview}, which is still an open problem.
\begin{figure}[htbp]
	\centering
	\includegraphics[width=\linewidth]{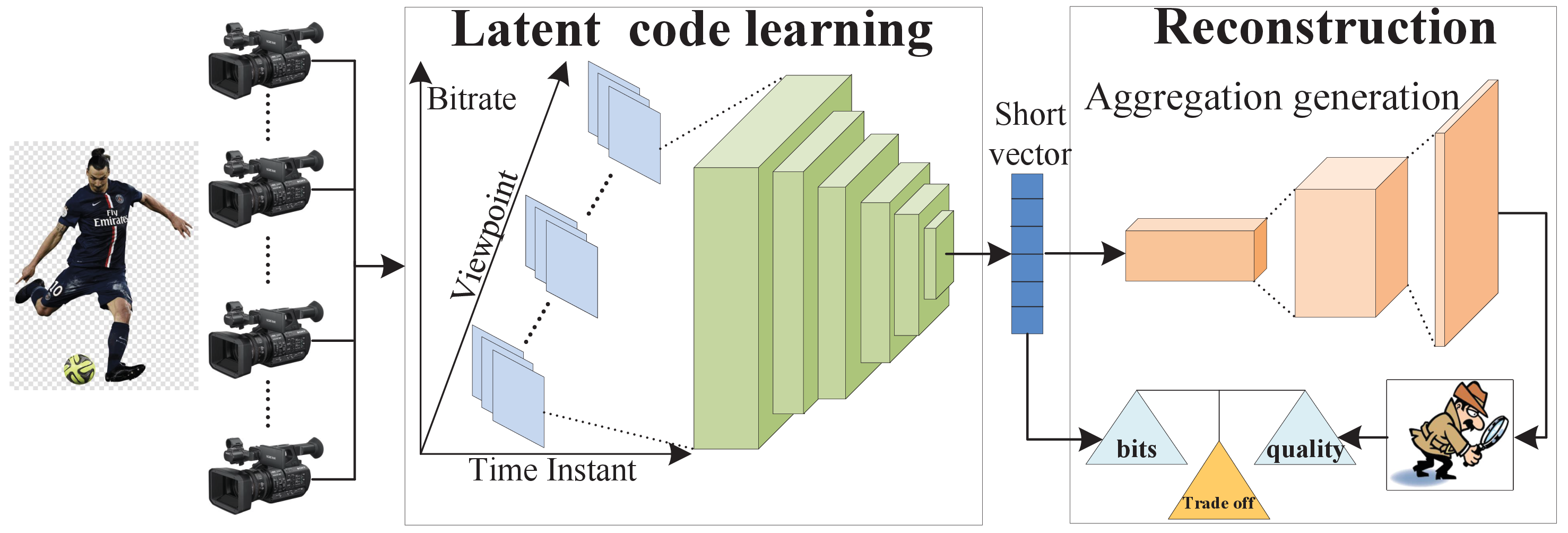}
	\caption{The methodology of our multi-view coding. By incorporating GAN latent learning, reconstruction and an RD optimization mechanism, our method significantly improves the reconstruction accuracy with a reduced bitrate.}
	\label{a}
\end{figure}

Traditional MVC methods utilize the hybrid coding framework to encode each viewpoint. To further reduce the output bitrates, a feasible approach is to skip intermediate viewpoints at encoder side and reconstruct them at decoder side. To this aim, small amount of information, which we call Side Information (SI), is introduced to extract the features of intermediate viewpoint for compensation of skipped information. Recently, the depth values of 3D scenes are often employed as SI to synthesize virtual viewpoint images by depth-image-based rendering \cite{tian2009view}. This is the currently  popular method called Multi-view plus Depth (MVD) \cite{muller20103}. However, the depth information cannot be accurately recovered due to the difficulty of obtaining and calculating the precise depth information\cite{2020-3d}. To improve the reconstruction quality, the low-resolution images are utilized as the SI in \cite{ hu2014multiresolution}. However, the compressed low-resolution images also lead to a high bitrate overhead, which limits its application. To address this issue, we propose to extract the spatio-temporal latent code of intermediate virtual viewpoint with Generative Adversarial Network (GAN) \cite{lipton2017precise}, which reduces bitrates overhead while maintaining the reconstruction quality of videos.

It is commonly known that neural networks can extract high-level semantic features, while GAN succeeds in generating images according to prior knowledge of sample datasets. Therefore, we deploy a GAN-based viewpoint reconstruction at decoder side. Meanwhile, an attempt is made to apply the latent vectors of GAN as SI. Previous studies suggest latent codes have contributed to restore original images \cite{chen2016infogan}. However, the conventional GAN approaches do not provide an inverse mapping  to project an image back into latent space. Recent years have witnessed a series of methods to extract latent code, including gradient descent \cite{creswell2018inverting} and adversarial feature learning \cite{donahue2016adversarial}.  Despite of these great efforts, they cannot be applied in our task due to the difficulty in establishing correlations between spatio-temporal and inter-view domains and lack of tradeoff between SI bitrates and reconstruction quality. As Epipolar Plane Image (EPI) method \cite{wu2018light} can aggregate information across viewpoints, we construct spatio-temporal EPI as input to the framework instead of the direct viewpoint to exploit correlations between spatio-temporal and inter-view domains. As depicted in Fig. \ref{a}, we address the bitrate and reconstruction quality trade-off of SI with a compact latent code and a bitrate optimization, which are capable of improving reconstruction accuracy with a reduced bitrates.

 Overall, we construct the spatio-temporal EPI as input to efficiently extract SI of the intermediate viewpoint and achieve better reconstruction quality. Furthermore, we use GAN to recover more details with high-level understanding across viewpoints. Finally, we use a hybrid cost function of reconstruction quality and bitrates to obtain the optimal SI while minimizing bitrates.
In summary, the contributions of this paper are as follows:
\begin{itemize}
\item We propose a multi-view video learning to exploiting the correlation between the spatio-temporal and inter-view domains to extract latent code as SI. The proposed method can effectively reduce the redundancy of SI.
\item We make the first attempt to introduce GAN coding network to reconstruct the intermediate viewpoint of MVC. The proposed method is able to accurately reconstruct intermediate viewpoint using latent code as SI.
\item We achieve a trade-off between reconstruct quality and SI bitrates by a bitrate optimization cost function. Experimental results show we achieve improved Rate-Distortion (RD) performance compared to the popular MVC methods.
\end{itemize}

\section{Related work}\label{section:TLBD}
This section focuses on current multi-view coding approaches and analyzes their shortcomings, including the MVC and deep-learning based MVC works.
\subsection{Multi-view video coding}
Multi-view video coding adds inter-view prediction to the standard of High Efficiency Video Coding (HEVC). It also introduces the concept of depth map, in which each viewpoint has an additional depth video. Therefore, we divide multi-view video coding into two categories: general multi-view video coding and depth map-based multi-view video coding.

\textbf{General multi-view video coding.}
MV-HEVC is the sate-of-the-art standard, which inspires many improvements on its modules. Hannuksela \emph{et al.} \cite{hannuksela2015overview} made a stage summary of multi-view extensions for HEVC and described the standard practices for multi-view video coding, which sets a milestone for the future work. Unlike the traditional RD model, Li \emph{et al.} \cite{li2020bit} proposed a multi-view bit allocation method based on the exact target bit relationship between base view and dependent view. To reduce coding complexity, Khan \emph{et al.} \cite{Khan2021} propose an Efficient Inter-Prediction Mode Decision (EIPMD) technique utilizes the Coding Unit (CU) splitting information of the base view to find its relation with the prediction Modes. Xu \emph{et al.} \cite{Xu2021} proposed a flexible complexity optimization framework that reduces encoder complexity according to external-defined constraints by minimizing the cost of alternative partitioning using a probability-driven approach known as APC. This method dynamically adjusts local candidate partitions to meet the target local complexity constraint. Li \emph{et al.} \cite{Li:SPIC21} proposed an RD optimization for the dependent viewpoint based on inter-viewpoint dependencies, and greatly improved its performance in MVC. In these general methods, the bitrates may increase sharply with the number of viewpoints, which is because the original video needs to be encoded in each viewpoint rather than the side information.

\textbf{Depth map-based multi-view video coding.}
Various coding methods have been proposed for depth map sequences from different aspects such as RD optimization, enhancement, bit allocation, and virtual view synthesis for depth maps. Müller \emph{et al.} \cite{muller20133d} improved motion compensation module to encode depth map sequences, and thus proposed an extended method for depth map HEVC based on inter-view prediction. By synthesizing the intermediate view with depth map and adjacent views, this method greatly saved bitrates and thus set a major milestone in the development of MVC. To address the problem of degraded quality at boundaries of synthesized viewpoints, Rahaman \emph{et al.} \cite{rahaman2017virtual} used Gaussian Mixture Models (GMM) to separate the foreground and fills holes in synthesized views. Besides, the amount of data transmitted can be further reduced by frame interpolation. Considering the application of depth map in intermediate view construction, a feasible method to improve MVC is to obtain accurate depth maps. Yang \emph{et al.} \cite{yang2018cross} recommended a cross-view multi-lateral filtering scheme, which enhances the quality of depth map using color and depth priors from adjacent views at different time-slots. To reduce the complexity in coding mode selection, Zhang \emph{et al.} \cite{zhang2020highly} proposed an efficient MVD scheme based on depth histogram projection and allowable depth distortion. Lin \emph{et al.} \cite{LinJR:TMM21} proposed to accelerate 3D-HEVC deep intra-frame coding using the characteristics of the human visual system. For the above methods, their qualities are limited by the quality of the depth map.

\subsection{Deep learning-based MVC}
Deep learning have been introduced into MVC with significantly improved performances. These works include deep learning-based MVC optimization and deep learning-based MVC post-processing.

\begin{figure*}[htbp]
	\centering
	\includegraphics[width=\linewidth]{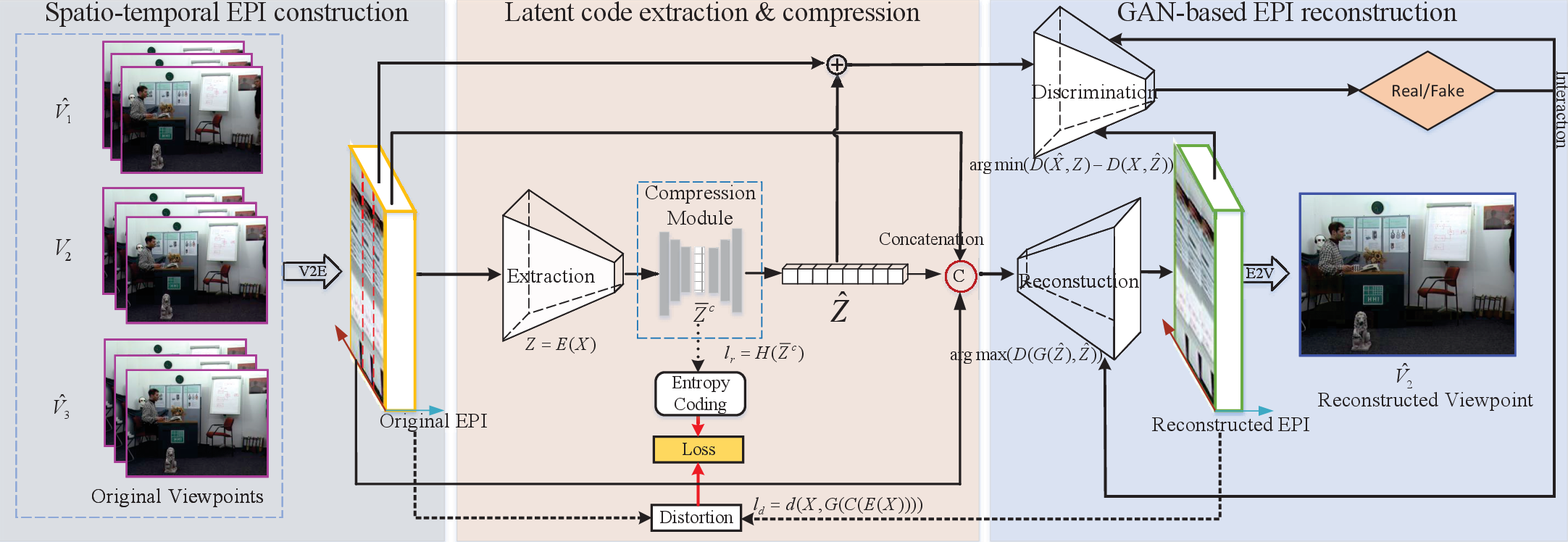}
	\caption{Framework of the proposed method, which consists of spatio-temporal EPI construction, latent
		code extraction \& compression and GAN-based EPI reconstruction. At the encoder side, we use the reconstructed left and right views as well as the original intermediate view to construct spatio-temporal EPI $X$. The brown arrows on $X$ indicate the viewpoint domain, while the blue arrows indicate the time domain. Then, $X$ is fed into the feature extraction network to extract the intermediate view latent code $Z$. The compression module compresses and reconstructs the latent code $Z$ with an auto-encoder and quantization.  After that, we reorganize EPI with the reconstructed latent code $\hat{Z}$, left view slice and right view slice by concatenation operation. Finally, GAN network reconstructs EPI $\hat{X}$ using the adversarial generation mechanism. In order to obtain the optimal RD performance, the latent extraction \& compression and GAN-based EPI reconstruction are jointly trained with a trade-off between reconstruction quality and bitrates.}
	\label{TC6}
\end{figure*}
\textbf{Deep learning-based MVC optimization.}
The deep learning-based MVC optimization approach introduces deep learning into specific modules of MVC framework. Lei \emph{et al.} \cite{Lei_2022} put forward a deep reference frame generation method for MVC. This method employed a Disparity-Aware reference frame Generation Network (DAG-Net) to transform the disparity relationship between different viewpoints and generate a more reliable reference frame. Lei \emph{et al.} \cite{Lei:TBC21} exploited spatial, temporal, and inter-view correlations and proposed a deep multi-domain prediction for 3D video coding. By employing CNNs to fuse multi-domain references, they achieved significant bitrates savings compared to 3D-HEVC. Peng \emph{et al.} \cite{Peng2023} proposed a multi-domain correlation learning module to recover the high frequency details of distorted frames by exploring multi-domain correlation. In addition, based on the block partition information generated in video coding, this method proposes a partition-constrained reconstruction module to better attenuate compression artifacts by designing partition losses. Chen \emph{et al.} \cite{LSVC} proposed the first end-to-end optimized stereo video coding framework to reduce redundancy by iteratively using motion estimation and disparity estimation for left and right views in feature space.

\textbf{Deep learning-based MVC post-processing.}
Deep learning-based methods are applied to the post-processing stage of the framework, which not only enhances quality of multi-view videos but also effectively removes compression artifacts. Recently, multi-frame quality enhancement approaches \cite{guan2019mfqe, Zhao_2021, Luo2022} have been proposed. They significantly reduce quality fluctuations between compressed video frames by locating peak-quality frames and enhancing low quality frames with adjacent high quality frames. He \emph{et al.} \cite{he2020mv} recommended a graph-neural-network-based compression artifacts reduction method, which reduces compression artifacts by fusing adjacent viewpoint messages and suppressing misleading information. Pan \emph{et al.} \cite{Pan:TCSVT21} proposed a Two-Stream Attention Network (TSAN)-based synthetic viewpoint quality enhancement method, which significantly improves the quality of synthetic viewpoints by extracting features at different scales using Multi-Scale Residual Attention Blocks (MSRAB) for enhancement. Zhang \emph{et al.}  \cite{Zhang:TCSVT22} modeled the elimination of temporal distortion as a perceptual video denoising problem and proposed a CNN-based synthetic viewpoint quality enhancement to reduce temporal flicker distortion while improving the perceptual quality of 3D synthetic viewpoint videos.

Inspired by these methods, we propose a method of multi-view latent code extraction and GAN-based multi-view video with spatio-temporal EPI reconstruction method to render new views by learning the least redundant latent code. The details of our proposed method are described as follows.

\section{Proposed Multi-View Latent Code Learning}\label{section:TPA}

The existing MVC methods usually extract and represent features in single viewpoint, {\it e.g.}, DCT coefficients, motion vectors and stereo depth maps, without regarding to simultaneous feature extraction and cross-viewpoint correlations. Capturing the correlation between multiple views simultaneously and leveraging the diversity in each view to achieve accurate original image reconstruction on the decoder side remains a critical and challenging issue.

\subsection{Problem formulation}
To facilitate the use of cross-viewpoint correlation, we employ the Epipolar Plane Image (EPI) method. On epipolar plane, an object projected into different viewpoints will appear on the same straight line of EPI. Using this approach, we group the corresponding data from different viewpoints into spatially adjacent locations in EPI, which enables us to exploit the inter-view correlation for subsequent processing. Then, the EPI is connected in temporal domain as the input of convolutional network to extract latent code. 

We model the latent code extraction of EPI:
\begin{equation}
Z = E(X),       
\end{equation}
where $X$ and $Z$ represent the original EPI and its high-level semantic features, respectively. $E$ represents a feature extraction network to extract the high-level semantic features of original EPI. To further reduce the bitrates of features, $Z$ will be compressed by using a compression network $C$ before being transmitted to the decoder side:
\begin{equation}
\hat Z = C(Z).       
\end{equation}
At the decoder side, we recover the EPI information with $\hat Z$ and a generator $G$.

Our task is to find an $E$ and its inverse operation $G$ to minimize the reconstruction error of EPI and the bitrates of latent code transmission:
\begin{equation}
\hat E,\hat G = \mathop {\arg \min }\limits_{E,G} \left\| {{X} - G(C(E(X)))} \right\|.      
\end{equation}

In this work, we employ CNN and GAN to perform latent code extraction \& compression and EPI reconstruction, respectively. The whole networks are co-trained in an end-to-end manner. As shown in Fig. \ref{TC6}, the whole framework consists of spatio-temporal EPI construction, latent code extraction \& compression and GAN-based EPI reconstruction, which are elaborated as follows.

\subsection{Spatio-temporal EPI construction} 
\begin{figure}[htbp]
	\centering
	\includegraphics[width=\linewidth]{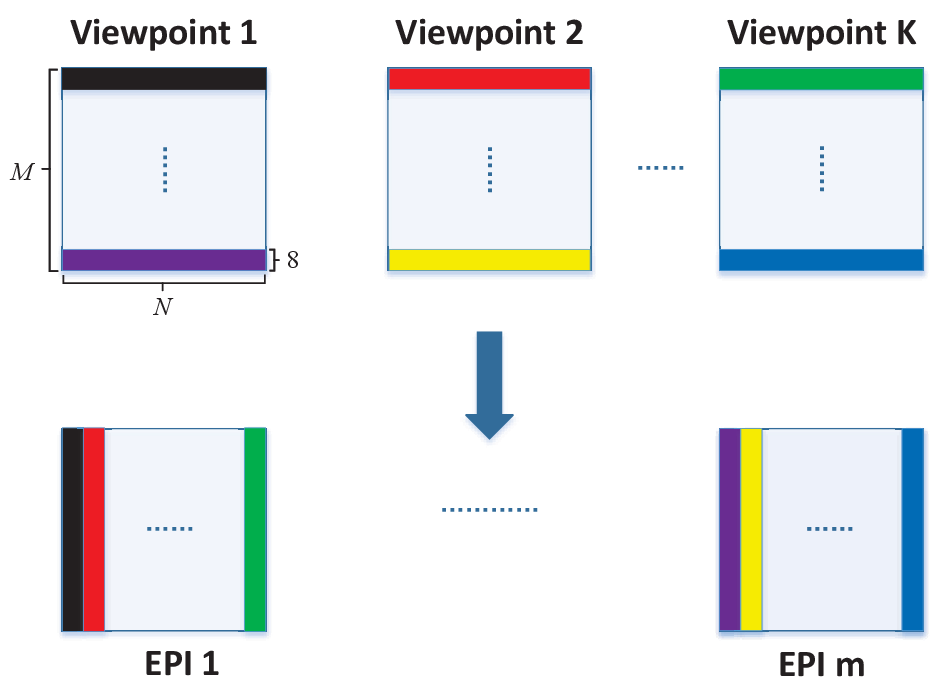}
	\caption{Construction of Spatial EPI.}
	\label{TAB2}
\end{figure}

Traditional EPI with pixel rows of images cannot well reflect the spatio-temporal correlations of multi-view images. In this work, we construct a spatio-temporal EPI with the following two steps. Firstly, the multi-view images are decomposed and reassembled based on their spatial locations. Let $M$, $N$ and $K$ denote the width, height and number of views of a multi-view video, respectively. As shown in Fig. \ref{TAB2}, images of each view are divided into $8\times N$ strips with all color channels, where the value 8 is chosen empirically\cite{bolles1987epipolar} and thus an image consists of $m=M/8$ strips. Then, all strips at the same spatial locations are grouped to formulate a spatial EPI with a dimension of $8K\times N\times3$. In total, there are $m$ spatial EPIs at a same time. Secondly, a spatio-temporal EPI is obtained by stacking $L$ successive spatial EPIs at time axis, in order to embed the temporal correlations. A spatio-temporal EPI is then with the dimension of $8K\times N\times 3L$. In particular, we set $L=3$ in this work. 

In the following, we refer a spatio-temporal EPI as ${\rm EPI}_j^t$, where $j=1,2,…,m$ and $t$ denotes the temporal index. The spatio-temporal EPI is then fed into convolutional layers to extract latent code.

\subsection{Latent code extraction \& compression}

\textbf{Latent code extraction.} As shown in Fig. \ref{TC6}, the latent code extraction step transforms the spatio-temproal EPI to a latent vector $Z$, which is critical to reconstruct the intermediate viewpoint. In order to obtain more accurate latent code with low bitrates, we use latent code extraction module to remove the redundant information at the encoder end. Through iterative training of the whole network, the module extracts latent code from the spatio-temporal EPI based on a trade-off between the quality of the reconstructed EPI and the bitrates. In this work, we achieve this with Fully Convolutional Network (FCN) due to its capacity to effectively learn data distribution with a compact feature size.

\begin{figure}[htbp]
	\centering
	\begin {minipage} [htbp] {0.215\textwidth}
	\centering
	\includegraphics [width=1.0\textwidth] {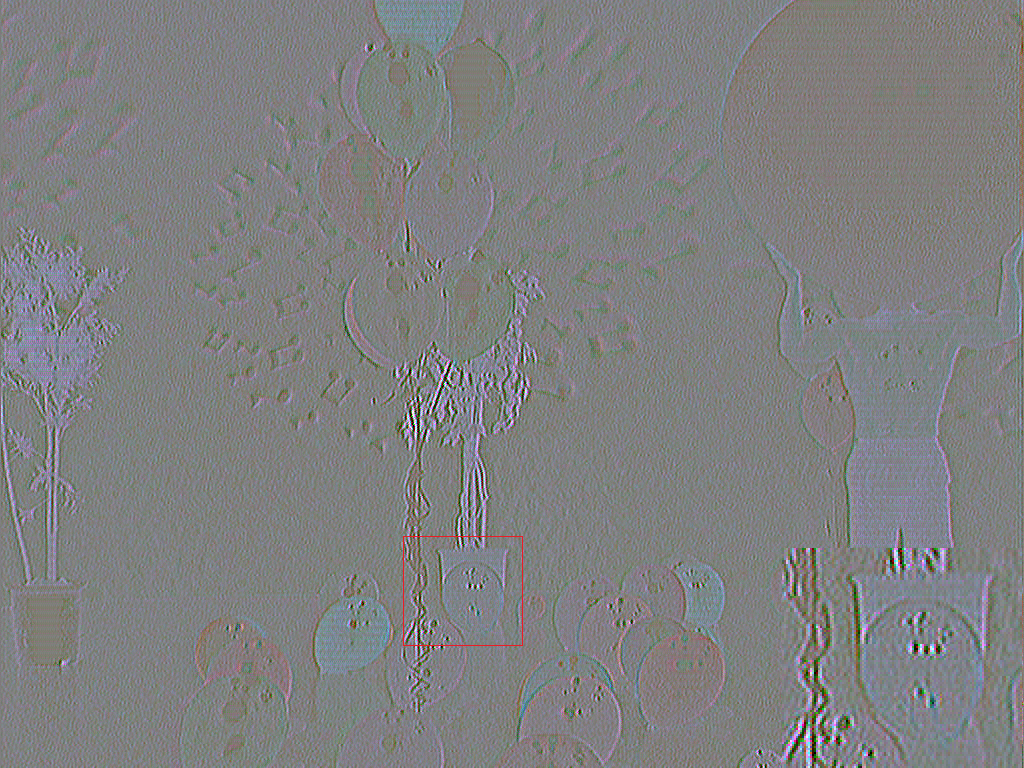}
	\centerline{(a)}
	\end {minipage}
	\begin {minipage} [htbp] {0.215\textwidth}
	\centering
	\includegraphics [width=1.0\textwidth] {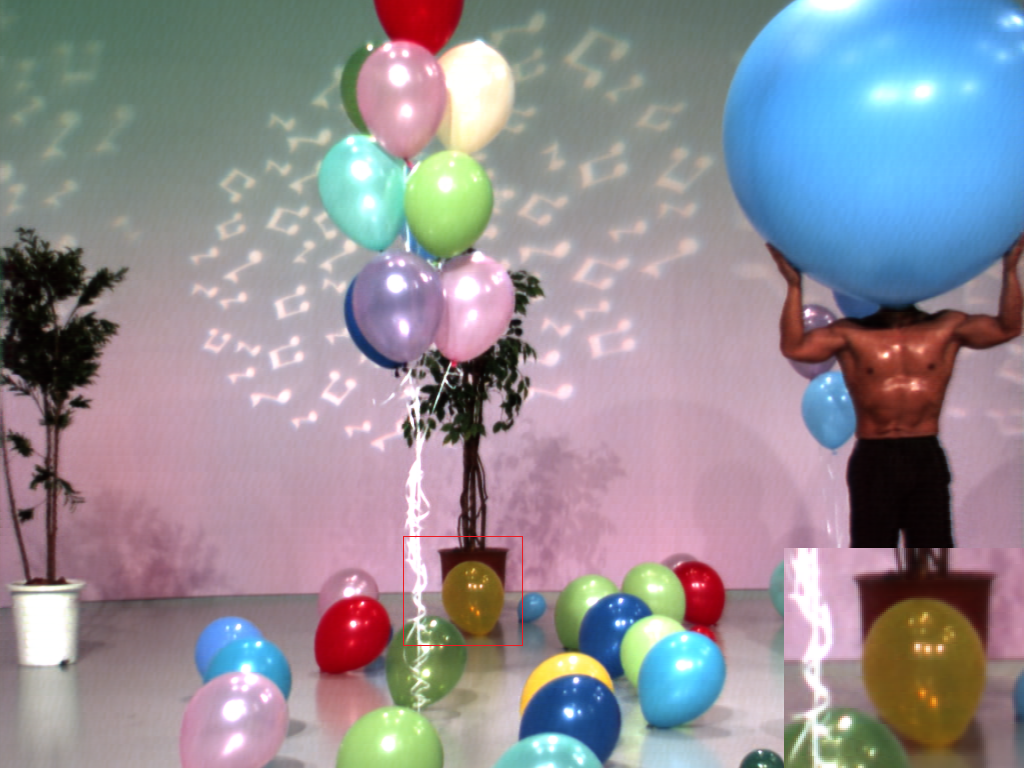}
	\centerline{(b)}
	\end {minipage}
	\caption{Visual examples from Balloons sequence. (a) Visualization of latent code. (b) Reconstruction with proposed method (115bits, 40.12dB).} 
	\label{vis}
\end{figure}

For the extraction module, its input and output are the spatio-temporal EPIs and latent code of intermediate viewpoint, respectively. The extraction network includes 1 convolutional layer, 4 residual blocks and another 2 convolutional layers. Each residual block includes 2 convolutional layers, 2 Bach Normalization (BN) layers, a Rectified Linear Unit (ReLU) function and an elementwise sum. The residual blocks are connected with skip connections and elementwise sum, in order to ensemble diverse feature information that benefits the visual quality of viewpoint reconstruction. As a result, the final output of extraction is with a dimension of $8$K$/3 \times N \times3L$ due to we only need the latent code of intermediate viewpoint. As shown in Fig.\ref{vis}, the latent code contains structural information of intermediate viewpoint, which can be used as a priori information of the GAN network to guide the generator $G$ to reconstruct the intermediate viewpoint.

\textbf{Latent code compression.} For the compression module $C$, we employ an auto-encoder with three downsampling layers and three upsampling layers that are implemented with convolutions and deconvolutions, respectively. The extracted latent code $Z$ will be compressed and quantized as a compact representation $\bar{Z}^c$, and then reconstructed as $\hat{Z}$.

\textbf{Entropy coding.} The quantized latent code $\bar{Z}^c$ will be transformed into bit-streams by performing entropy coding. In this work, we employ the hyperprior entropy model presented in \cite{lu2019dvc} for an accurate bitrates estimation. During the training process, the calculated bitrate is used as a part of the loss function.

\subsection{GAN-based EPI reconstruction}
By using a compact representation of EPI features, we provide multi-view video at a lower bitrate. At the receiver side, we reconstruct the generated EPI $\hat X$ from $\hat{Z}$, $\hat X=G(\hat{Z})$, where $G$ denotes the function of reconstruction process. To ensure that the reconstructed EPI $\hat X$ is as close as possible to the original EPI $X$, we refer to the GAN framework to introduce the discrimination function $D$. Through the interaction between $D$ and $G$, the EPI generated by $G$ can increasingly approximate the original EPI $X$. As shown in Fig. \ref{TC6}, the EPI reconstruction is composed of two modules: reconstruction and discrimination, as well as an interaction mechanism.

The reconstruction module uses a neural network with convolution as the generator. We consider the following two distributions: Joint probability density function in latent code extraction $p(Z,X) = p(X)p(Z|X)$; Joint probability density function in reconstruction $p(X,\hat{Z}) = p(\hat{Z})p(X|\hat{Z})$. In these distributions, $p(X)$ is the prior probability function of the original EPI, $p(Z)$ is the probability density function of the latent code and $p(\hat{Z})$ is the probability density function of the compressed latent code. In the latent code extraction \& compression process, the extraction network $E$ maps the original EPI $X$ to the latent code $Z$: $Z=E(X)$ and the compression network $C$ compresses $Z$ to $\hat{Z}$, while in the reconstruction process, the generator network $G$ maps the samples of the compressed prior $p(\hat{Z})$ to the input space $X = G(\hat{Z})$. To accurately reconstruct the EPI, it is necessary to make the conditional probability $p(X|\hat{Z})$ coincide with the prior probability $p(X)$ as much as possible.

The discrimination module determines
whether the input EPI is the original EPI by a classification network. To better discriminate whether the input image is the original EPI, the training goal is to make the value of $D(X,E(X))$ as large as possible and the value of $D(G(\hat{Z}),Z)$ as small as possible. Unlike the GAN network which only discriminates the input image, our discriminator network needs to discriminate both the image and the latent code.

The interaction mechanism uses the discrimination results to guide the generator to reconstruct images closer to the original EPI, and also to navigate the discriminator to better identify differences between the generator’s output and the original EPI. To this end, we design this mechanism with an adversarial game in which the discriminator and the generator are trained alternately. The discriminator is trained to distinguish between sample pairs from the encoder $(X,\hat Z=C(E(X)))$ and sample pairs from the generator $(\hat X=G(\hat{Z}),Z)$, both of which satisfy the joint probability distribution $p(X,\hat{Z})$ or $p(Z,X)$. The generators are trained to fool the discriminators obtained from the previous training. At this point, the discriminant value $D(X,\hat Z)$  is as large as possible.
\begin{algorithm}[t]
	\caption{Minibatch Adam Stochastic Optimization descent training of  three  networks of $E$, $C$, $G$, and $D$.}
	\label{MVLL}
	Pretrain and initialize networks $E$, $C$ and $G$, and then train the network with the data sets samples $X = \{ {X^{(1)}},...,{X^{(k)}}\} $ of EPIs as follows:
	\begin{algorithmic}
		\FOR {epoch in range (0, epochs)}
		\STATE Update learning rate with exponential decay.\\
		\FOR {index in range (0, iterations, batchsize)}
		\STATE Sample minibatch of $k$ examples from data of EPIs as input.\\
		\STATE Update the discriminator network by ascending its stochastic gradient:\\
		\STATE ${\nabla _D}\frac{1}{k}\sum\limits_{i = 1}^k {\left[ {\log D({X^i},Z) + \log (1 - D(G(\hat Z),\hat Z))} \right]}$
		\STATE Update the generator network by descending its stochastic gradient:\\
		\STATE ${\nabla _G}\frac{1}{k}\sum\limits_{i = 1}^k {\left[ {\log (1 - D(G(\hat Z),\hat Z)) + d({X^i},G(\hat Z))} \right]}$\\
		\STATE Update the extraction and compression network by descending its stochastic gradient:\\
		\STATE ${\nabla _E}\frac{1}{k}\sum\limits_{i = 1}^k \{ {V_{GAN}}({X^i}) + \lambda [d(X,G(C(E({X^i}))))]$\\
		\quad\quad\quad\quad\quad\quad${+ H(\bar{Z}^c))\}}$\\
		\ENDFOR
		\ENDFOR
	\end{algorithmic}
\end{algorithm}

\subsection{The objective function }
The latent code extraction \& compression and GAN-based EPI reconstruction are co-trained within a network flow. A joint objective function is thereby designed and utilized to optimize both the bitrates of latent code and the visual quality of reconstructed videos. As discussed above, the spatio-temporal EPI $X$ is extracted as latent code $Z$ by convolutional network $E$, compressed to $\hat Z$ by compression module $C$ and finally reconstructed as $\hat X$ by generator $G$. A discriminator $D$ is deployed to identify the reconstruction performance. Inspired by GAN, the objective function of whole network can be expressed as:
\begin{equation} 
	\begin{array}{l}
		\quad\quad\mathop{\min}\limits_{E,G}\mathop{\max}\limits_{D} \{E_{X{\sim}p_X}[\log D(X,Z)]\\\quad\quad+E_{Z{\sim}p_Z}[\log(1-D(G(\hat{Z}),\hat{Z}))]\}+\lambda {l_d} + {l_r},   
	\end{array}
\end{equation}

\noindent where $p_X$ and $p_Z$ denote the distributions of spatio-temporal EPI $X$ and latent code $Z$ respectively. $l_d$ denotes the distortion loss and $l_r$ denote the numbers of bits used to encode latent code. $\lambda$ is a hyper parameter used to control the rate-distortion trade-off.


We calculate the distortion and bitrates losses based on distance and entropy, respectively. The distortion loss is calculated between the original and reconstructed spatio-temproal EPIs, 
\begin{equation}
l_d = d(X, G(C(E(X)))),    
\end{equation}
where the distance $d(\cdot)$ is obtained as a combination of pixel-domain Mean Squared Error (MSE) and feature-domain VGG loss,
\begin{equation} 
\begin{array}{l}
d(x,y) = l_{MSE}(x,y)+ l_{VGG}(x,y)\\
=\frac{{1}}{wh}\sum\limits_{i = 1}^w\sum\limits_{j = 1}^h(x_{i,j}-y_{i,j})^2~
+\frac{{1}}{wh}\sum\limits_{i = 1}^w\sum\limits_{j = 1}^h(\phi(x_{i,j})-\phi(y_{i,j}))^2.
\end{array}
\end{equation}
Here $w$ and $h$ represent the dimension of EPI. $\phi(\cdot)$ denotes the operation of the VGG network to extract the feature map.\par
As depicted in Section 3.3, we can use the entropy model to accurately estimate the bitrates of the compressed and quantized latent code {$\bar{Z}^c$}:
\begin{equation}
	l_r = H(\bar{Z}^c).
\end{equation}

\subsection{The overall algorithm}
With the loss function defined in Section 3.5, we are able to train the whole network consisting of the extraction module $E$, the compression module $C$, the generator $G$ and the discriminator $D$. The whole training process is summarized in the Algorithm\ref{MVLL}.

\section{Experiments}\label{section:ER}
\subsection{Experiment setup}
\textbf{Datasets.} We train our method, which is named as MVLL for the sake of simplicity, with 5 typical multi-view sequences including {\it Scence\_Door\_Flowers} (1024 $\times$ 768), {\it Scence\_Leaving\_Laptop} (1024 $\times$ 768), {\it Scence\_Outdoor} (1024 $\times$ 768), {\it Champagne\_Tower} (1280 $\times$ 960) and {\it Dog} (1280 $\times$ 960). Each of them is with 5 views. They are further split into EPI images, according to the steps shown in Section 3.2. We selected 12,800 EPIs from each produced EPI sequence, and a total of 64,000 EPIs were used for training. To examine the performance of our method, we test on the Common Test Conditions (CTC) of MVC \cite{muller2014common}. The details are listed in Table \ref{tablea}. We selected 30 frames of video from each sequence for the EPI construction and put them to the test. \par
\begin{table}[htbp]
	\centering
	\caption{The video sequences for testing.}
	\label{tablea}
	\begin{tabular}{c|ccc}
		\hline
		Sequence & Resolution & Viewpoints & Frame Rate \\
		\hline
		Balloons & 1024$\times$768 & 1, 2, 3 & 30 \\
		Bookarrival & 1024$\times$768 & 6, 7, 8 & 16.67\\
		Kendo & 1024$\times$768 & 1, 2, 3 & 30\\
		Lovebord1 & 1024$\times$768 & 4, 5, 6 & 30\\
		Newspaper & 1024$\times$768 & 2, 3, 4 & 30\\
		Pantomime & 1280$\times$960 & 37, 38, 39 & 30\\
		Dancer & 1920$\times$1088 & 1, 5, 9 & 25\\
		Poznan\_Street & 1920$\times$1088 & 3, 4, 5 & 25\\
		\hline
		
	\end{tabular}
\end{table}
\textbf{Baseline codecs.} There are four other methods implemented for performance comparison. The state-of-the-art multi-view video coding standard, MV-HEVC \cite{hannuksela2015overview}. The multi-view video plus depth standard method, MVD \cite{tech2015overview}, uses the depth maps as the auxiliary information. The deep-learning based video coding method, DCVC \cite{Li:NIPS21}, used conditional coding to replace the traditional predictive coding paradigm. The deep-learning based multi-view video enhancement method, TSAN \cite{Pan:TCSVT21}, used a dual-stream attention network to improve the synthesized view quality. To make a fair comparison, for MV-HEVC, we use the HTM-16.3 software with {\it baseCfg\_3view.cfg} configuration setting and set the QP to \{30, 35, 40, 45\}, while for MVD, we use the HTM-16.3 software and VSRS3.5 view synthesis software with {\it baseCfg\_2view+depth.cfg} configuration and setting the QP pairs of $(QP_t,QP_d)$ for texture and depth videos, i.e. (30, 39), (35, 42), (40, 45) and (45, 48). For the DCVC method, we use the HTM16.3 software  with {\it baseCfg\_2view.cfg} configuration setting to encode the left and right views, and use the DCVC method with their pre-trained models to compress the intermediate view. For the TSAN method, we use their pre-trained models to perform quality enhancement on the intermediate view synthesized by the MVD method.\par
\textbf{Metrics.} The popular video coding criteria are employed to evaluate and compare all the above four methods. These criteria include the PSNR-aware measurements (rate-PSNR curves, the BDPSNR and BDBR \cite{bjontegaard2001calculation}) and the SSIM-aware measurements (rate-SSIM curves, the ADSSIM and ADBR \cite{zhao2015ssim}).\par
\textbf{Implementation details.} In our implementation, we use the HTM-16.3 software with {\it baseCfg\_2view.cfg} configuration setting to encode the left and right views, and compress the intermediate view using our proposed method. We train four models with different $\lambda$ values ($\lambda$ = 512, 1024, 2048, 4096). The other parameters for training are set as follows: epochs=50, iterations=64000, batchsize=1. 
\begin{figure*}[htbp]\centering
	\begin {minipage} [htbp] {0.24\textwidth}
	\centering
	\includegraphics [width=1.0\textwidth] {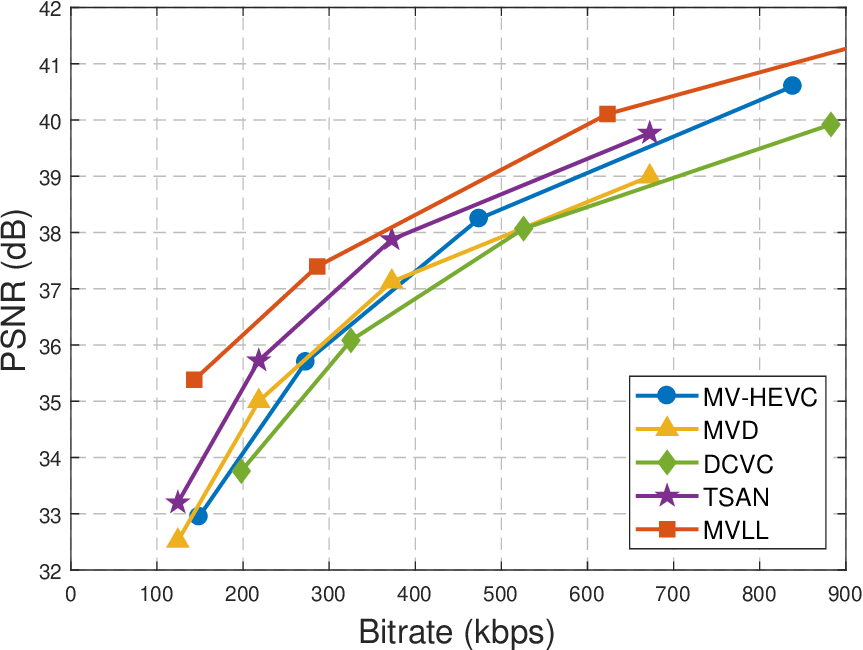}
	\centerline{(a) Balloons}
	\end {minipage}
	\begin {minipage} [htbp] {0.24\textwidth}
	\centering
	\includegraphics [width=1.0\textwidth] {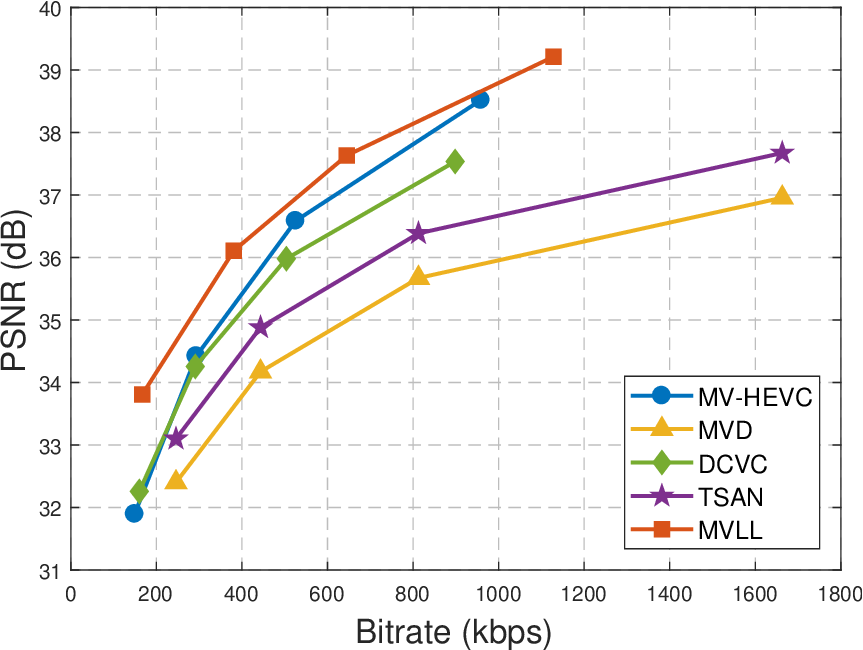}
	\centerline{ (b) Bookarrival}
	\end {minipage}
	\begin {minipage} [htbp] {0.24\textwidth}
	\centering
	\includegraphics [width=1.0\textwidth] {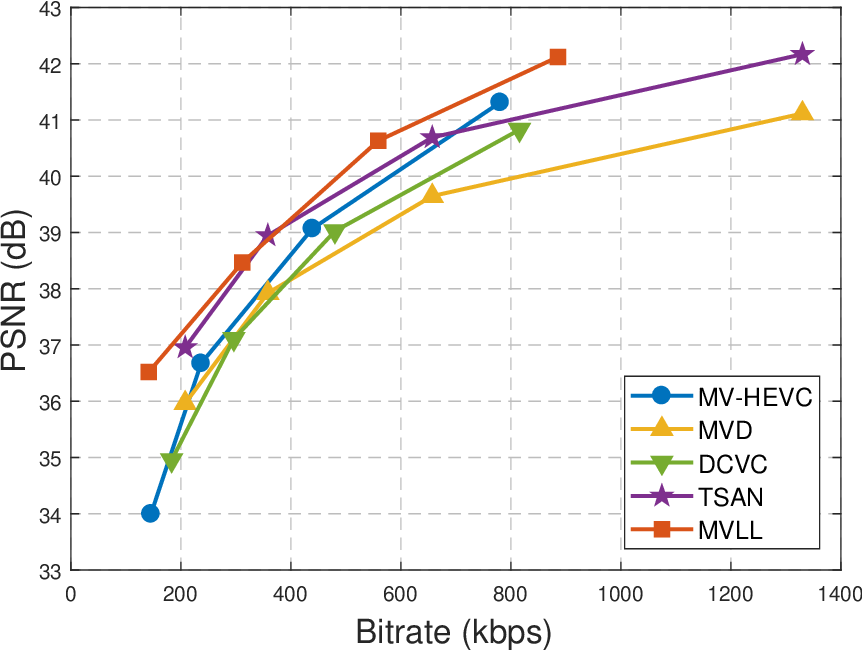}
	\centerline{ (c) Kendo}
	\end {minipage}
	\begin {minipage} [htbp] {0.24\textwidth}
	\centering
	\includegraphics [width=1.0\textwidth] {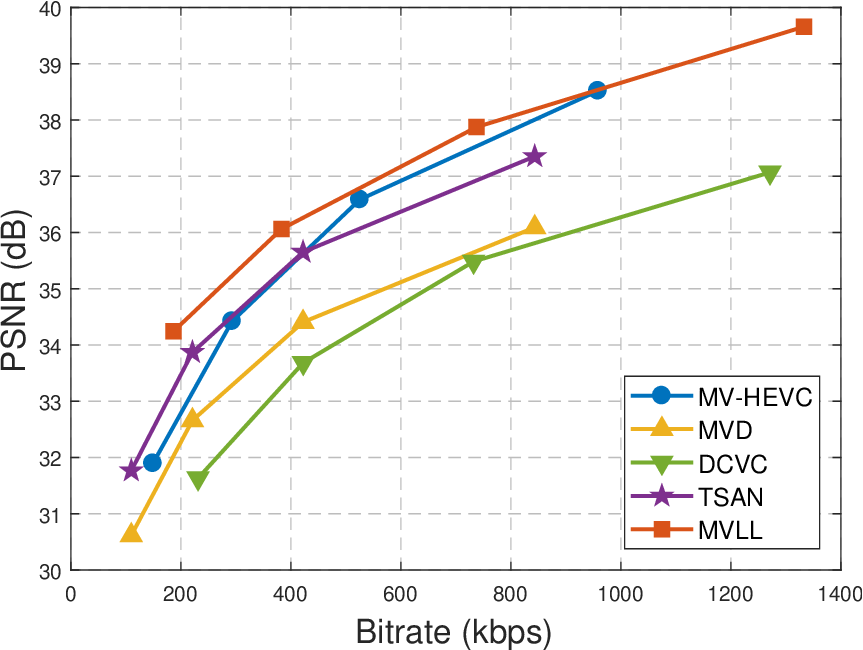}
	\centerline{ (d) Lovebird1}
	\end {minipage}
	\begin {minipage} [htbp] {0.24\textwidth}
	\centering
	\includegraphics [width=1.0\textwidth] {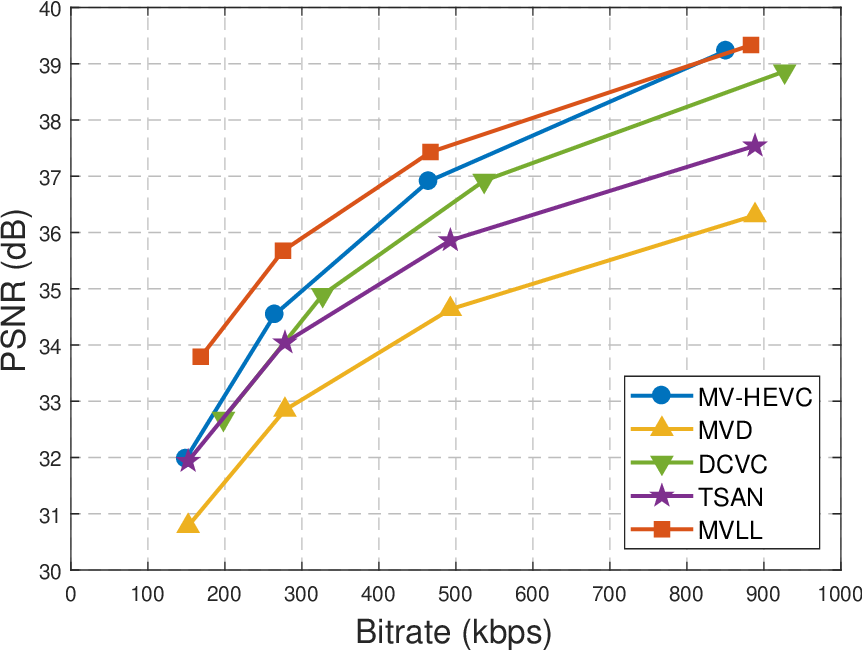}
	\centerline{(e) Newspaper}
	\end {minipage}
	\begin {minipage} [htbp] {0.24\textwidth}
	\centering
	\includegraphics [width=1.0\textwidth] {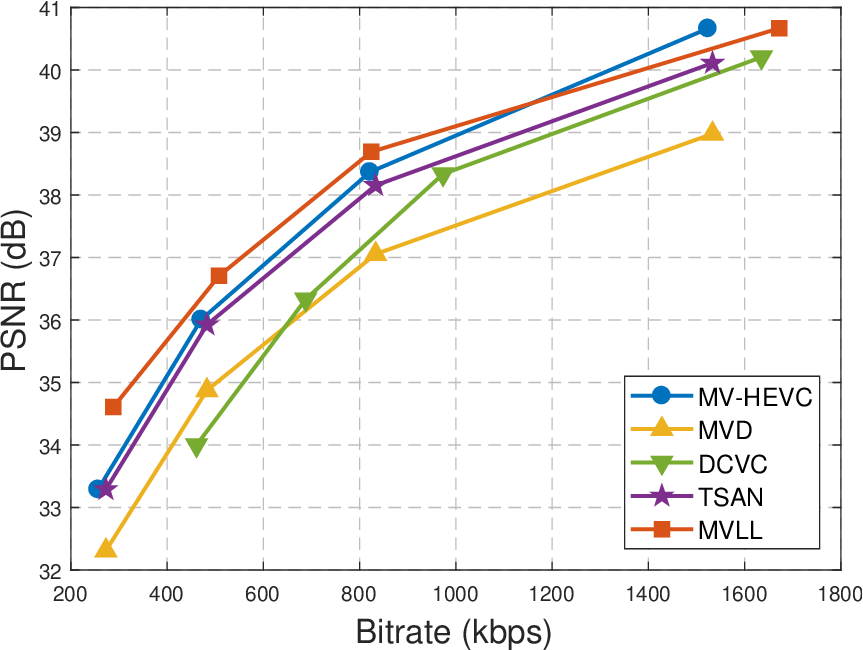}
	\centerline{(f) Pantomime}
	\end {minipage}
	\begin {minipage} [htbp] {0.24\textwidth}
	\centering
	\includegraphics [width=1.0\textwidth] {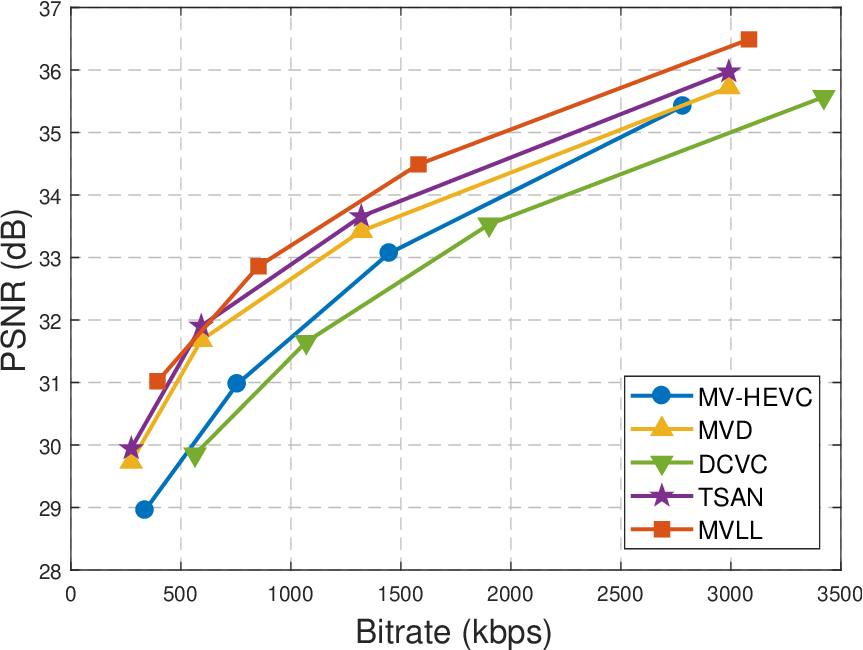}
	\centerline{(g) Dancer}
	\end {minipage}
	\begin {minipage} [htbp] {0.24\textwidth}
	\centering
	\includegraphics [width=1.0\textwidth] {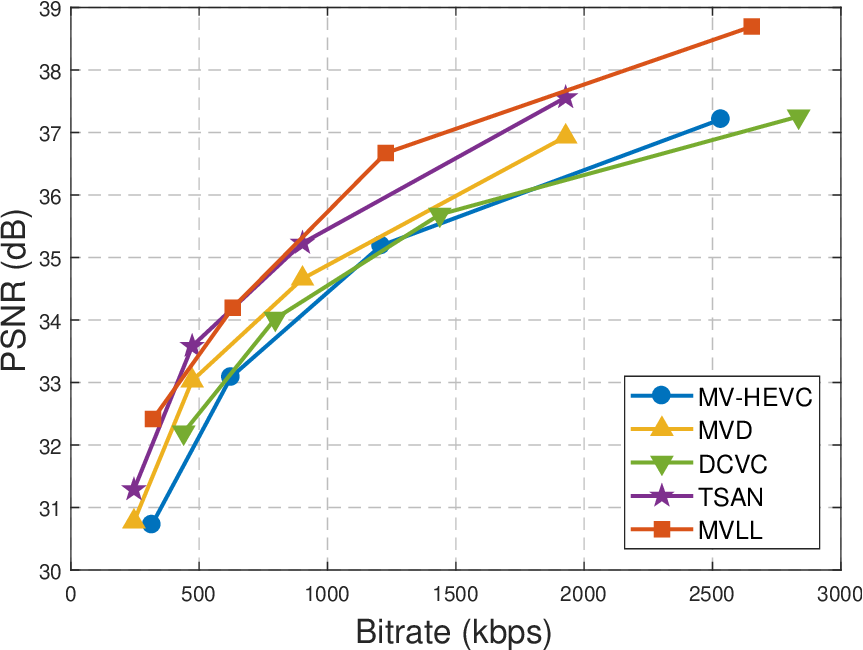}
	\centerline{(h) PoznanStreet}
	\end {minipage}
	\caption{Rate-PSNR curves of all compared methods.} 
	\label{pa}
\end{figure*}

\begin{figure*}[htbp]\centering
	\begin {minipage} [htbp] {0.24\textwidth}
	\centering
	\includegraphics [width=1.0\textwidth] {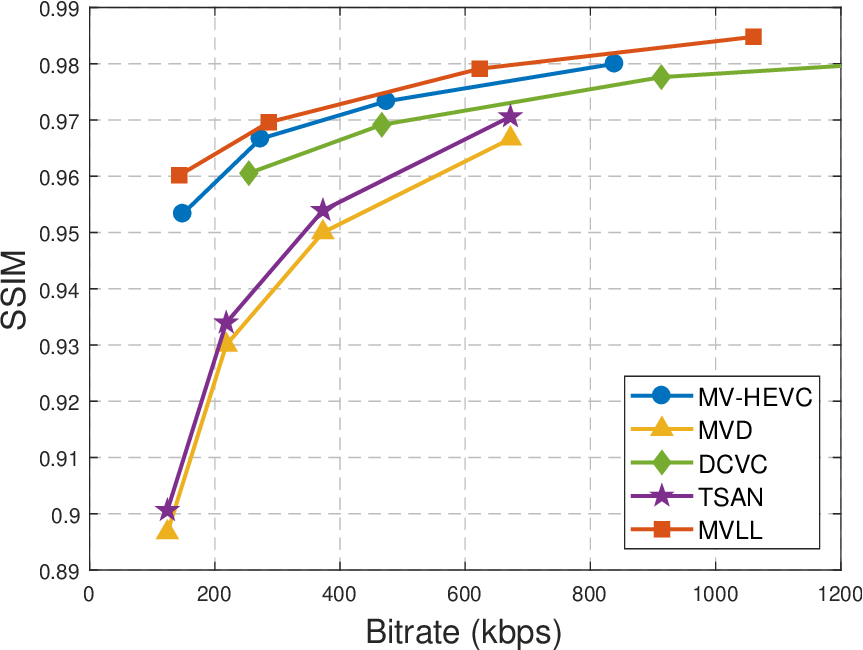}
	\centerline{(a) Balloons}
	\end {minipage}
	\begin {minipage} [htbp] {0.24\textwidth}
	\centering
	\includegraphics [width=1.0\textwidth] {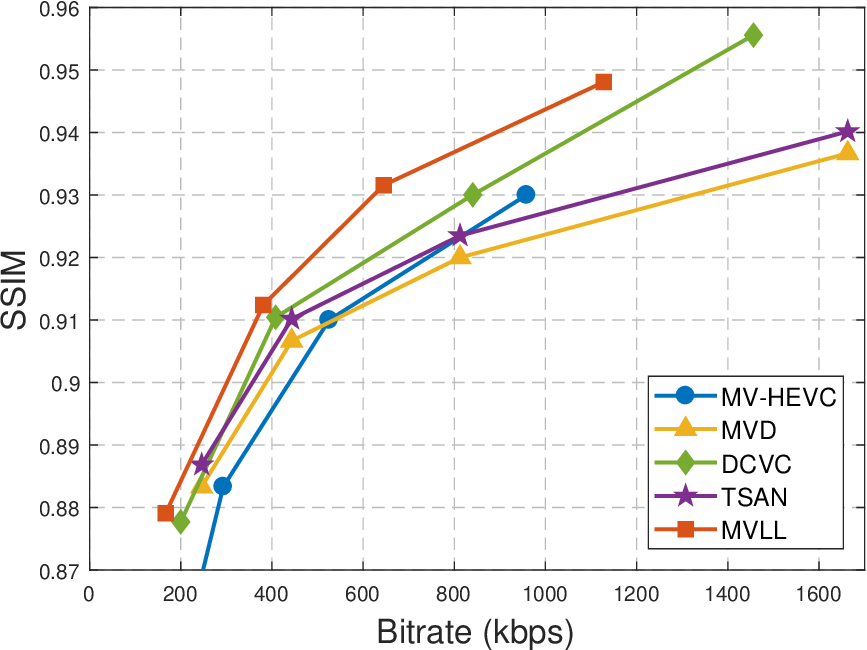}
	\centerline{ (b) Bookarrival}
	\end {minipage}
	\begin {minipage} [htbp] {0.24\textwidth}
	\centering
	\includegraphics [width=1.0\textwidth] {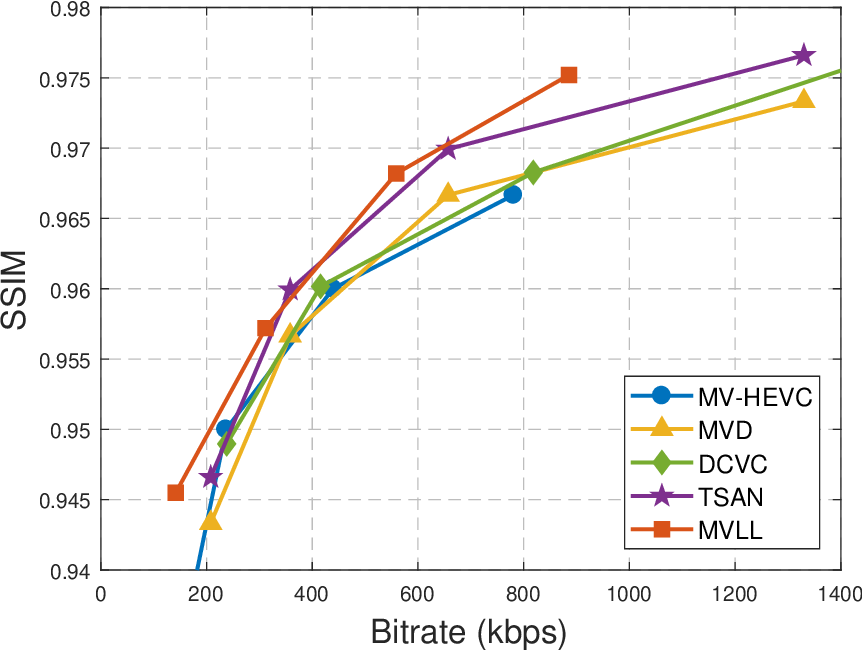}
	\centerline{ (c) Kendo}
	\end {minipage}
	\begin {minipage} [htbp] {0.24\textwidth}
	\centering
	\includegraphics [width=1.0\textwidth] {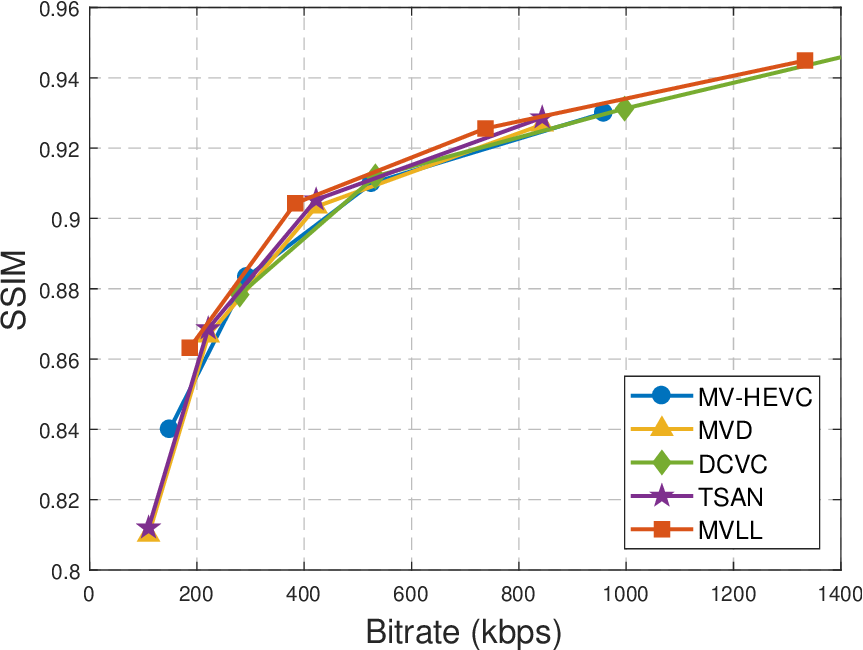}
	\centerline{ (d) Lovebird1}
	\end {minipage}
	\begin {minipage} [htbp] {0.24\textwidth}
	\centering
	\includegraphics [width=1.0\textwidth] {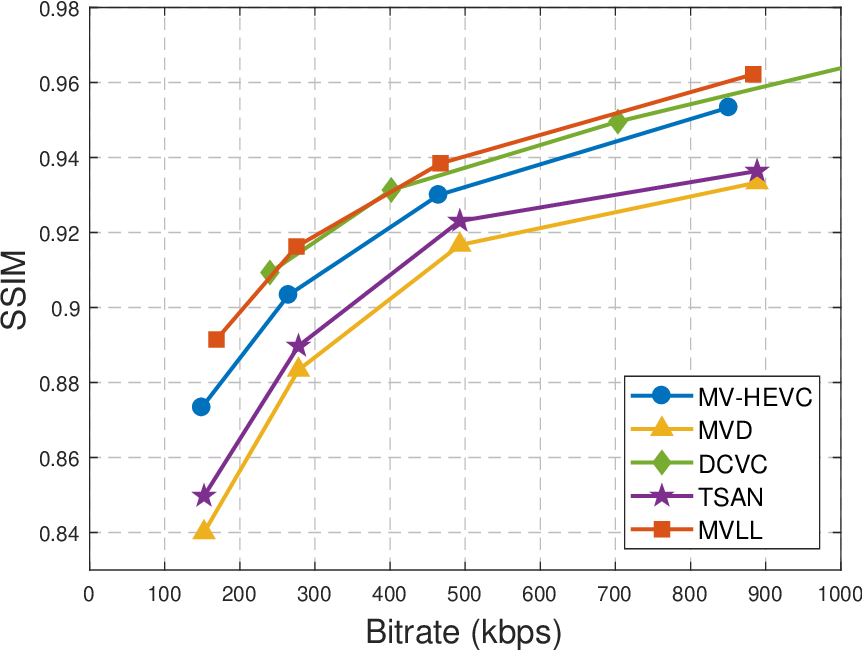}
	\centerline{(e) Newspaper}
	\end {minipage}
	\begin {minipage} [htbp] {0.24\textwidth}
	\centering
	\includegraphics [width=1.0\textwidth] {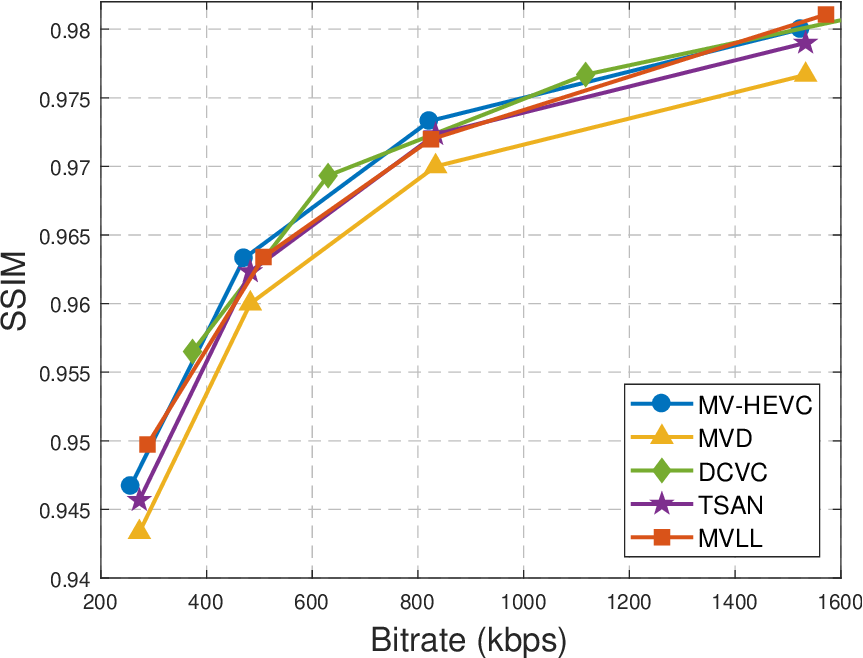}
	\centerline{(f) Pantomime}
	\end {minipage}
	\begin {minipage} [htbp] {0.24\textwidth}
	\centering
	\includegraphics [width=1.0\textwidth] {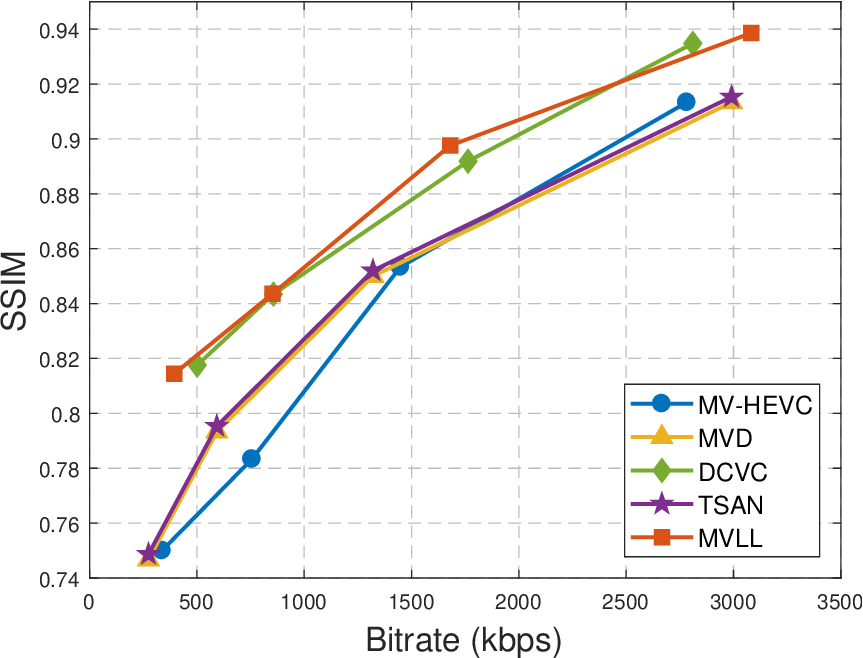}
	\centerline{(g) Dancer}
	\end {minipage}
	\begin {minipage} [htbp] {0.24\textwidth}
	\centering
	\includegraphics [width=1.0\textwidth] {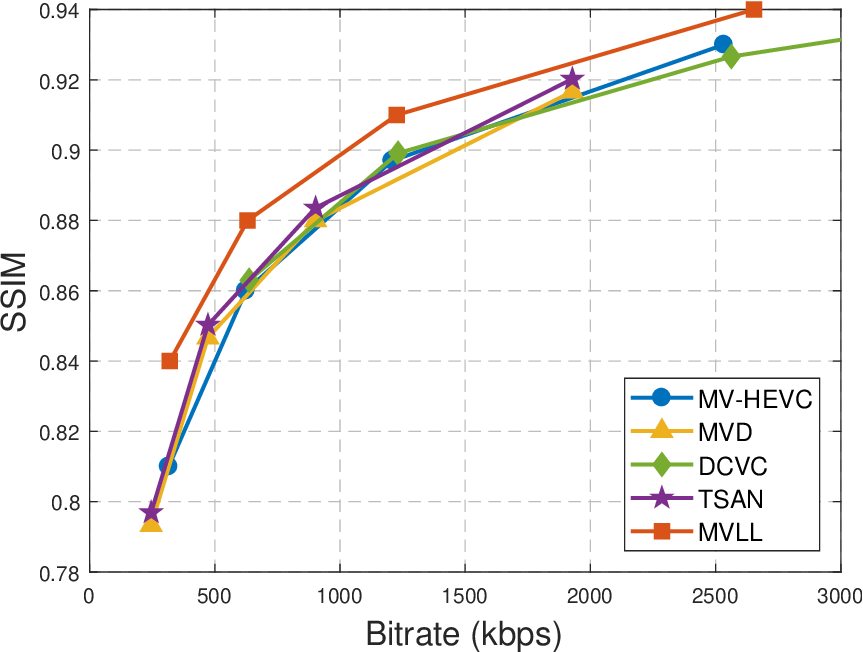}
	\centerline{(h) PoznanStreet}
	\end {minipage}
	\caption{Rate-SSIM curves of all compared methods.} 
	\label{pb}
\end{figure*}
\begin{table*}[htbp]
\renewcommand{\arraystretch}{1.2}
\centering
\caption{The encoder improvements evaluated by PSNR.}
\resizebox{\textwidth}{20mm}{
\label{tableb}
\begin{tabular}{c|c|c|c|c|c|c|c|c|c|c}
\hline
\quad&\quad&\emph{Balloons}&\emph{Bookarrival}&\emph{Kendo}&\emph{Lovebird1}&\emph{Newspaper}&\emph{Pantomime}&\emph{Dancer}&\emph{PoznanStreet}&Average\\
\hline
\multirow{2}{*}{BDBR($\%$)}&\bfseries MV-HEVC &-23.7886&-17.1885&-17.9240&-13.5281&-6.9199&-32.7008
& -68.3621& -36.1006& -20.4410\\ \cline{2-11}
                           &\bfseries MVD&-40.1095&-71.2063&-39.8545&-57.4701&-65.5207&-38.2222
&-17.6076&-25.7746&-44.1407\\ \cline{2-11}
										&\bfseries DCVC&-36.4340&-31.4592&--28.8073&-59.2842&-26.7557&-27.6334& -41.8136& -38.0574&-36.2806 \\
						\cline{2-11}
										&\bfseries TSAN&-20.5295&-56.2097&-10.9087&-28.1795&-43.8484&-14.1459& -9.9345& -7.1604&-23.8646\\
\hline
\multirow{2}{*}{BDPSNR}&\bfseries MV-HEVC&1.1203&0.6381&0.7341&0.6249&0.4609&-0.0827& 1.3435& 0.9945& 0.7290\\ 
				\cline{2-11}
                       &\bfseries MVD&1.5722&2.6093&1.4409&2.1399&2.9809&1.4293&0.4759&0.7446&1.6741\\ \cline{2-11}
									 &\bfseries DCVC&1.8429&1.0744&1.3624&2.6531&1.2096&1.5557& 1.7062& 1.2880& 1.5804\\
						\cline{2-11}
									 &\bfseries TSAN &0.8242&1.9015&0.4191&0.9056&1.7666&0.3368& 0.2444& 0.1538& 0.8190\\
\hline
 \end{tabular}}
\end{table*}
\begin{table*}[htbp]
\renewcommand{\arraystretch}{1.2}
\centering
\caption{The encoder improvements evaluated by SSIM.}
\resizebox{\textwidth}{20mm}{
\label{tablec}
\begin{tabular}{c|c|c|c|c|c|c|c|c|c|c}
\hline
\quad&\quad&\emph{Balloons}&\emph{Bookarrival}&\emph{Kendo}&\emph{Lovebird1}&\emph{Newspaper}&\emph{Pantomime}&\emph{Dancer}&\emph{PoznanStreet}&Average\\
\hline
\multirow{2}{*}{ADBR($\%$)}&\bfseries MV-HEVC&-18.9829&-33.6628&-19.5974&-14.5925&-18.8311&5.8075&-25.1623&-25.5533&-18.8219\\ \cline{2-11}
                           &\bfseries MVD&-69.4709&-37.1942&-24.0581&-11.8847&-45.6205&-13.5243&-30.2773 &-28.6590 &-32.5861 \\ \cline{2-11}
											&\bfseries DCVC&-41.1030&-19.7054&-24.2316&-11.8847&-45.6205&-13.5243&-30.2773& -25.4233&-16.6478\\
										\cline{2-11}
										&\bfseries TSAN&-61.8296&-31.1066&-71.1281&-7.5817&-37.7603&-1.9382&-28.6742& -23.1616&-32.8975\\
\hline
\multirow{2}{*}{ADSSIM}&\bfseries MV-HEVC&0.0030&0.0178&0.0044&0.0066&0.0095&0.0010&0.0426&0.0163&0.0124\\ \cline{2-11}
                       &\bfseries MVD&0.0288&0.0144&0.0041&0.0056&0.0303&0.0028&0.0289 &0.0166 &0.0505\\ \cline{2-11}
									&\bfseries DCVC&0.0064&0.0063&0.0036&0.0044&0.0020&-0.0009&0.0034&0.0122&0.0047\\
									\cline{2-11}
									&\bfseries TSAN&0.0248&0.0108&0.0008&0.0036&0.0242&0.0004&0.0268&0.0130&0.0131\\
\hline
\end{tabular}}
\end{table*}
\subsection{Experimental results}
Fig. \ref{pa} shows the rate-PSNR curves of all compared methods: MV-HEVC, MVD, DCVC, TSAN and MVLL. In each curve, the four data samples correspond to the four Qp settings; while in each data sample, the rate-PSNR performances are obtained as the averaged results of all views. Fig. \ref{pb} shows the rate-SSIM curves of these methods, where the SSIM index is considered is more consistent with human vision system than PSNR.

From Figs. \ref{pa}-\ref{pb} we can get several conclusions. {\textbf Firstly}, all methods achieve good performances in terms of R-Q curves, by retaining a high compression ratios at the acceptable visual quality. Generally speaking, they are still superior to the original encoder without optimizations. {\textbf Secondly}, the MV-HEVC methods achieve superior performance than MVD in most sequences, which may be due to the extra bitrates for depth maps in MVD. In case of multi-view plus depth coding, the MVD method is still preferred. {\textbf Thirdly}, the DCVC achieves acceptable performance but still inferior to our MVLL. This method was designed for general video coding and thus do not exploit the inter-view correlations between multi-view videos. {\textbf Fourthly}, the TSAN methods enhances the quality of synthetic view of the MVD method, thus achieves superior performance than MVD in all sequences. {\textbf Fifthly}, our MVLL method outperforms the compared methods in most sequences, which is contributed to its GAN-based inter-view prediction. 

Tables \ref{tableb} and \ref{tablec} presents the quantitative comparisons of R-Q performances, where the efficiency of our MVLL is evaluated with MV-HEVC, MVD, DCVC and TSAN as the benchmarks. The terms BDBR and ADBR indicate the average bitrates savings at the same PSNR and SSIM, respectively. The terms BDPSNR and ADSSIM indicate the average quality increments at the same bitrates, respectively. From the tables, we can see the quantitative comparison results are consistent with those in Figs. \ref{pa}-\ref{pb}. Compared with the other methods, our MVLL reduces 20.44\%, 44.14\%, 36.28\% and 23.86\% bitrates at the same PSNR, or 18.82\%, 32.59\%, 16.65\% and 32.90\% bitrates at the same SSIM. It also significantly improves the video quality in terms of PSNR and SSIM at the same bitrates. These results demonstrate the efficiency and effectiveness of our proposed method.

As for the computational complexity, we use the balloons sequence to evaluate our encoder time on the machine with a single 2080TI GPU (11GB Memory). The coding time of our proposed method is 0.204s per frame, which is faster than the traditional hand-crafted MVC standard MV-HEVC or 3D-HEVC.



\subsection{Ablation study}
The main contributions of the proposed method consist of two parts: the spatio-temporal EPI structure, and the GAN based autoencoder network structures. Therefore, to validate their effectiveness, the ablation experiments are performed.\par
\begin{figure}[htbp]
	\centering
	\includegraphics [width=0.7\linewidth]{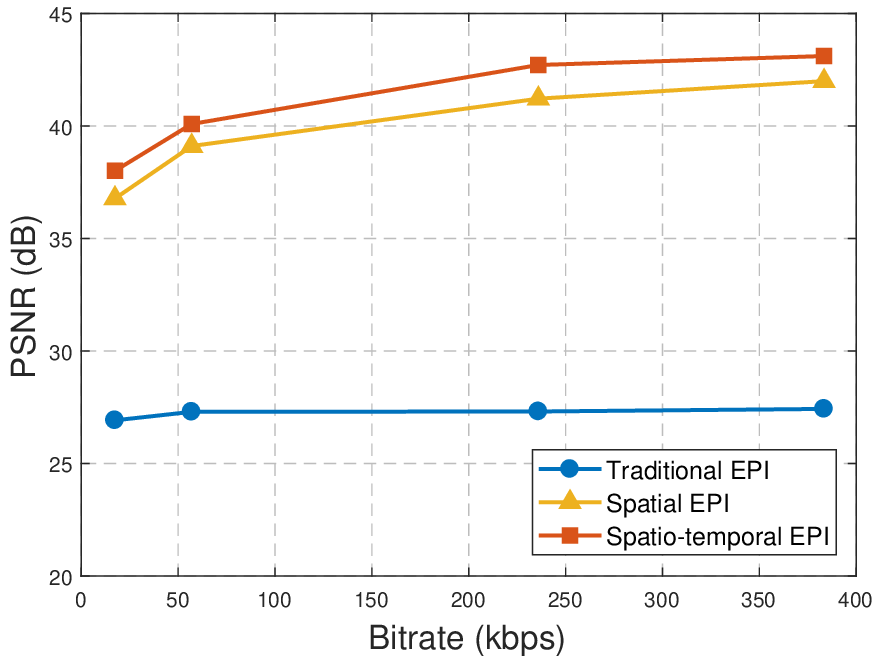}
	\caption{Comparison of the rate distortion performance of traditional EPI, spatial EPI and spatio-temporal EPI, where the results are obtained by averaging 30 frames of the Kendo test sequence.} 
	\label{xiaorong1}
\end{figure}

\textbf{Contribution of spatio-temporal EPI structure:} To analyze the effect of the structure of EPI on the results of MVLL, we compare the algorithm performance of three different input structures: traditional EPI, spatial EPI and spatio-temporal EPI. In the traditional EPI structure, we set the width of the column occupied by each viewpoint in the EPI to 1. In the spatial EPI, the column width occupied by each viewpoint in the EPI is 8. And in the spatio-temporal EPI, we stack the time-domain information of the 3 spatial EPIs on 9 different channels in the intermediate viewpoint. With other experimental conditions being consistent, Fig. \ref{xiaorong1} represents the rate distortion results of the algorithmic based on the three EPI structures, where the horizontal axis is the bitrates of the latent code in the intermediate viewpoint. Since applying a different EPI structure requires retraining the neural network model, we use a training data set of size 8000 to reduce the computational complexity. As can be seen in Fig. \ref{xiaorong1}, with the increase of the bitrates of the latent code, the image quality of the reconstructed image results from the algorithm of the traditional EPI structure grows slowly. However, the image quality growths of the results from both the algorithm of the spatial EPI and the spatio-temporal EPI structures are more obvious. Furthermore, the image quality of the spatio-temporal EPI structure is superior to the performance of the spatial EPI. Therefore, we choose to use the structure of spatio-temporal EPI in this paper.\par

\begin{figure}[htbp]
	\centering
	\includegraphics [width=0.7\linewidth]{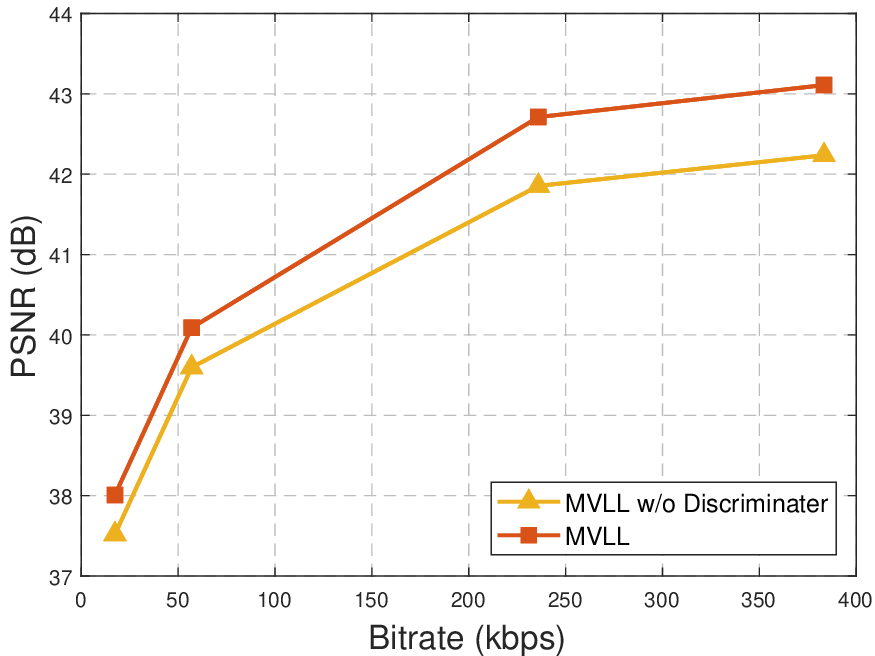}
	\caption{Comparison of RD performance for different network architectures, where the results are obtained by averaging 30 frames of the Kendo test sequence.} 
	\label{xiaorong2}
\end{figure}

\textbf{Comparison of GAN and autoencoder network structures:} To analyze the contribution of the GAN network structure, we compared the performance of both GAN and autoencoder network structures. Since the difference between these two network structures is whether they have a discriminator network or not, we implement the two different algorithms by adding or subtracting the discriminator network from the original network structure. With all other experimental conditions being equal, Fig. \ref{xiaorong2} represents the rate distortion performance results of the multi-view video coding algorithm based on latent code learning for two network structures, GAN and autoencoder network. We can see that the rate distortion performance of the algorithm with discriminator network structure is higher. Therefore, in this paper we use a GAN structure with discriminator network for multi-view video coding.

\section{Conclusion}\label{section:Conclusion}

Nowadays, there exists a bottleneck to further compress video streams with the traditional hybrid model. Researchers have been contributing to deep-learning-based video compression, where the motion prediction/compensation, RD optimization or entropy coding is realized by deep network. In this paper, we made the first attempt to combine deep GAN model with multi-view video coding. We utilized the latent code of GAN as SI in an RD-optimal manner. The latent code is generated with a deep network and further utilized to reconstruct the intermediate views, thereby saving the streaming bitrates of multi-view videos. Experimental results show a significant performance gain over the state-of-the-art schemes using depth map as SI. We hope this work can provide an innovative methodology to deep-learning-based multi-view video coding.

\bibliographystyle{ACM-Reference-Format}
\bibliography{ref}

\end{document}